\documentclass[
preprint,
showpacs,preprintnumbers,
showkeys,
 amsmath,amssymb,
 aps,
 prl,
]{revtex4-1}

\usepackage{graphicx}% Include figure files
\usepackage{dcolumn}% Align table columns on decimal point
\usepackage{bm}% bold math
\begin{document}

\preprint{Draft}

\title{Stimulated excitation of an optical cavity \\
by a multi-bunch electron beam\\
 via coherent diffraction radiation process}
%\thanks{A footnote to the article title}

\author{Yosuke Honda}
 \email{yosuke@post.kek.jp}
\author{Miho Shimada}%
\author{Alexander Aryshev}%
\author{Ryukou Kato}%
\author{Tsukasa Miyajima}%
\author{Takashi Obina}%
\author{Ryota Takai}%
\author{Takashi Uchiyama}%
\author{Naoto Yamamoto}%
\affiliation{%
High Energy Accelerator Research Organization (KEK), 1-1 Oho, Tsukuba, Ibaraki, Japan
}%

\date{\today}% It is always \today, today,

\begin{abstract}
With a low emittance and short-bunch electron beam
at a high repetition rate
realized by a superconducting linac,
stimulated excitation of an optical cavity 
at the terahertz spectrum range has been shown.
The electron beam passed through small holes
in the cavity mirrors
without being destroyed.
A sharp resonance structure
which indicated wide-band stimulated emission via 
coherent diffraction radiation was observed
while scanning the round-trip length of the cavity.
\end{abstract}

% Free-electron lasers 41.60.Cr
% Transition radiation 41.60.Dk
% Resonators, cavities, amplifiers, arrays and rings 42.60.Da
% Infrared , submillimeter wave, microwave and radiowave sources 07.57.Hm
% Beam dynamics; collective effects and instabilities 29.27.Bd

%Use showkeys class option if keyword
%display desired

\maketitle

%\tableofcontents

\paragraph{Introduction}

Light sources
have played important roles in progress of science in various fields.
The technologies of light sources are most immature in the terahertz range,
which is usually defined from 0.3 to 3 THz.
Coherent emission of electromagnetic radiation 
from a short-bunch electron beam in an electron linac
can be used as a terahertz radiation source.
With an energy-recovery linac (ERL) scheme \cite{erlreview}
which realizes a high average current beam in a linac layout
by recycling the beam energy,
a high power terahertz source can be realized 
\cite{jlabcsr, krafft_source, novo-fel}.
Conventionally,
the coherent synchrotron radiation (CSR)
has been considered as the radiation mechanism,
including the layout of a bending magnet or an undulator.
On the other hand, 
various other mechanisms can also be used,
including transition radiation and diffraction radiation.
One advantage of these over the synchrotron radiation
is the simple geometry located in a straight pass.
Another feature is the spatial mode of the radiation,
which are radiated in a higher-order transverse mode with radial polarization.
Coherent transition radiation (CTR)
is emitted when a short-bunch electron beam hits a metal target.
CTR is widely used for electron beam diagnostics of accelerators
\cite{ctr-kung,ctr-murokh}.
However,
it cannot be used as a high power source
because it destroys the electron beam.
Coherent diffraction radiation (CDR) \cite{dr-potylitsyn}
is a similar radiation mechanism as CTR.
Radiation is emitted when an electron beam 
non-destructively passes near the target.

Here,
we consider an optical cavity system 
which stacks the coherent radiation emitted
in the cavity,
in other words, the beam excites the optical cavity.
When the cavity is excited at resonance,
an electron bunch coherently emits radiation
in the electromagnetic field that already exists in the cavity.
This results in extracting more radiated power from the electron bunch
than a simple setup that is not based on a cavity.
This mechanism is called stimulated radiation.
Such a system has been proposed 
for applications in a high-flux X-ray production \cite{shimada_csr, shimada_csr2}.

The principle of stimulated radiation
has been tested by CTR \cite{lihn}
and CSR \cite{shibata_broadband,shibata_ctr,shibata_temporal}
in destructive layouts.
In these experiments,
by measuring the radiation power
while scanning the round-trip length of the optical cavities,
one can observe sharp resonance peaks due to stimulated radiation.
A test in CDR layout has been performed
with a limited number of bunches \cite{aryshev_lucx}.

The advantage of CDR layout is
its availability in a high power accelerator.
A superconducting linac that can produce 
a short-bunch beam 
at high repetition rate
in a continuous operational mode fits this advantage.
We report an experiment performed at a modern superconducting linac
constructed as a test facility of ERL \cite{cerlconstruction}.
The low emittance beam produced by the photocathode injector 
can realize the CDR layout which requires the beam pass through a small aperture.
The sub-ps short bunch beam
generated in bunch compression mode \cite{cerlbunch}
emits coherent radiation in the terahertz range.
We present a result showing evidence of stimulated radiation
produced in an optical cavity 
by a multi-bunch beam passing through the cavity.
To the best of our knowledge, 
this is the first experiment to clearly show
a stimulated radiation signal
in the terahertz range
with a CDR layout.

%%%%%%%%%%%%%
\paragraph{Principle}

\begin{figure}[b]
\includegraphics[width=0.9\linewidth, height=0.45 \linewidth]{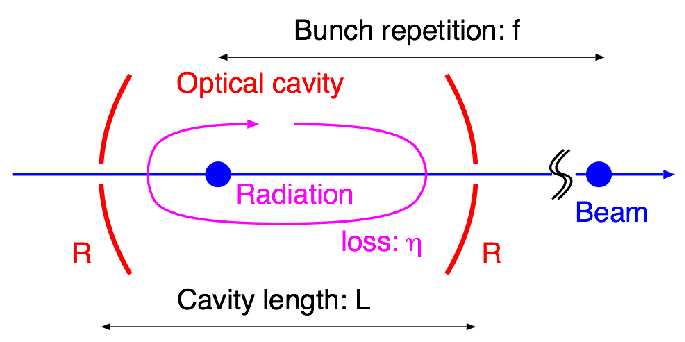}% Here is how to import EPS art
\caption{\label{fig:principle} Layout of the system. 
The beam passes through the center holes of the optical cavity.
}
\end{figure}

We consider a situation shown in Figure \ref{fig:principle}.
An optical cavity formed by two concave mirrors
with a small hole in the center
is installed in a straight pass of an electron beam.
The cavity length $L$ is the distance between the mirrors.
And it is designed so as to match the round-trip time to the bunch repetition rate $f$.
For simplicity,
we limit our discussion to a symmetric case,
i.e., the two mirrors have the same curvature
and the system has cylindrical symmetry.

%%%%%%%%%%%%
%%%%%%%%%%%%

The eigenmodes of the cavity
couple with the beam via the electric field along the beam trajectory.
Since the fundamental transverse mode of the cavity
does not have a longitudinal field at the center,
it cannot be excited in this layout.
On the other hand, 
the odd higher-order transverse modes can be excited.
For example, 
the transverse electric field of the 1-st-order mode is written as follows \cite{siegman}:
\begin{equation}
E^r = \frac{A}{w(z)} \frac{r}{w(z)}
\exp \left( - \frac{r^2}{w^2(z)}\right)
\cdot
\exp[i(\omega t - kz) + \phi (z)]
\end{equation}
where $z$ is the beam direction and
$r$ is the transverse distance from the beam axis.
The transverse beam size is described as $w(z) = w_0 \sqrt{1+(z/z_0)^2}$,
where $w_0$ and $z_0$ are the beam waist and the Rayleigh length, respectively.
They are related by $z_0=\pi w_0^2/\lambda$,
where  $\lambda$ is the radiation wavelength,
$k = 2 \pi/\lambda$, and $\omega/k=c$.
$\phi(z)$ is known as the Gouy phase, which depends on the order of the transverse mode,
where $\phi(z) = 2 \tan ^{-1} (z/z_0)$ for the 1-st-order mode.
The longitudinal field of the mode can be obtained from the general relation
\cite{laseracc}, 
$ik E^z = \partial E^r / \partial r $.
The beam moves at the speed of light $c$ along the axis
and encounters a longitudinal field of
\begin{equation}
E^z = - \frac{A}{k w^2(z)} \exp[i \phi(z)] \quad .
\end{equation}
The excitation energy in the cavity mode 
can be calculated from
the well-known relation describing the coupling between the beam and cavity:
\cite{handbook};
\begin{equation}
U^{exc} = \frac{q^2}{4 U} \left| \int E^z dz \right|^2 \quad ,
\end{equation}
where $q$ is the bunch charge.
$U$ is the energy stored in the cavity corresponding to the electric field $E^r$.
The integration should be done along the beam trajectory within the cavity.
Since the cavity length 
is fixed based on the bunch repetition rate,
the only free parameter 
is the curvature radius of the mirrors $R$.
The eigenmode size $w_0$ can be changed by 
selecting the appropriate value of $R$.
From the calculation above,
the maximum beam coupling is realized at $R=L$,
which is the so-called confocal cavity design.

In the case of multi-bunch excitation,
the signal of each bunch stacks
as a coherent amplitude addition.
The amplitude after the $n$-th bunch becomes
$
v_n = v_1 \sum_{m=1}^n \left(\sqrt{1-\eta} e^{i \theta} \right)^m
$
, where $\eta$ is the power loss and
$\theta$ is the phase shift over a single round-trip, respectively.
$\theta$ can be changed by finely changing $L$ in a scale within the wavelength.
The power enhancement gain after an infinite number of bunches
can be obtained as
\begin{equation}
G = \frac{|v_{\infty}|^2}{|v_1|^2} 
= \frac{1}{2-\eta - 2 \sqrt{1- \eta} \cos \theta} \quad.
\end{equation}
The enhancement gain at resonance is
$G=4/\eta^2$.
High radiation power is extracted 
at a lower cavity loss.

The optical cavity has many longitudinal modes
that correspond to frequencies that are integer multiples of the round-trip frequency.
In general, the Gouy phase $\phi$ for each longitudinal mode
is different,
and hence, the resonance condition is different for each mode.
Shift of the resonance condition for the $i$-th longitudinal mode
is given as
\begin{equation}
\Delta \theta^{(i)} = 2 \pi
\left( i - \frac{4}{\pi} \tan^{-1}\sqrt{\frac{L/R}{2-L/R}}\right)
\quad .
\end{equation}
In the special confocal cavity design,
$\Delta \theta^{(i)}$ becomes an integer multiple of $2\pi$ for all $i$,
resulting in all longitudinal modes being excited simultaneously.
Under a similar consideration,
it can be shown that 
all odd ordered transverse modes are also excited under the same condition.
This mechanism can be described as
a picture of the carrier-envelope phase (CEP) of the pulse
traveling back-and-forth in the optical cavity.
The confocal cavity is a special case with zero-CEP shift,
and hence, it can coherently add broad spectral signals in a multi-bunch beam.

\begin{figure}[b]
\includegraphics[width=0.9\linewidth, height=0.45\linewidth]{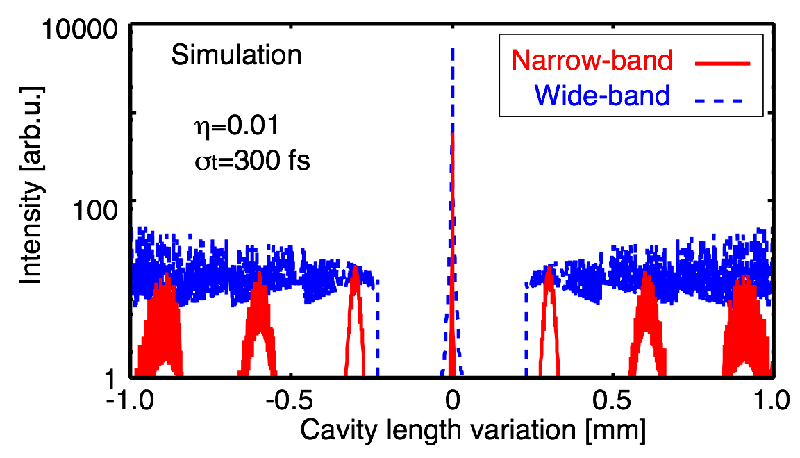}% Here is how to import EPS art
\caption{\label{fig:simulation} Simulation of the cavity length scan.
Excited power in the cavity is plotted as a function of cavity length variation
with respect to the perfect synchronization condition.
The wide-band case 
refers to sum of the signals for all longitudinal modes.
The narrow-band case
refers to the weighted sum of the longitudinal modes,
assuming the effect of a realistic band-pass filter with
0.5 THz center frequency and 10 \% bandwidth (full-width half-maximum)
at the detector.
}
\end{figure}

In the experimental case,
we finely scan $L$ while measuring the excited power in the cavity
under a given beam repetition rate.
Figure \ref{fig:simulation} shows a calculation example. 
The cavity loss is $\eta = 0.01$.
This calculation was performed 
by including the effects of the bunch spectrum, assuming an RMS bunch length of 300 fs.
A sharp resonance peak is observed
when the resonance conditions are satisfied for all modes,
which we call perfect synchronization.
The width of the resonance is referred to as the finesse $F$,
which is defined as the ratio between the half-wavelength 
and the fullwidth-half-maximum of the resonance peak.
The finesse is determined only by the cavity loss as $F\sim2\pi/\eta$.

%%%%%%%%%%%%
%%%%%%%%%%%%

\paragraph{Experimental Setup}

We performed an experiment at cERL \cite{cerlconstruction} in KEK.
This experiment was performed at the burst operation of energy non-recovery mode.
In the burst operation,
the electron beam emission at the gun was time-gated
by a photo-cathode laser system,
while all the RF systems in the accelerator cavities were operated in CW.
The beam condition in this experiment was as follows;
The beam energy 
was 17.8 MeV,
the bunch charge was set to be 1.2 pC,
and the  macro-pulse duration was 1 $\mu$s,
which contains 1300 bunches at 1.3 GHz repetition.
The beam was operated in bunch compression mode \cite{cerlbunch}.
A bunch length shorter than 300 fs was realized in the straight section
where our experimental setup was installed.
The normalized beam emittance in the straight section was measured 
to be 1.4 mm$\cdot$mrad.
The electron beam optics were adjusted to focus the beam at the location
using two quadrupole magnets
placed at 1.58 m and 4.78 m upstream of the cavity.
The RMS beam size in the horizontal and vertical planes
were measured to be 
250 $\mu$m and 60 $\mu$m, respectively.

\begin{figure}[b]
\includegraphics[width=0.9\linewidth, height=0.5\linewidth]{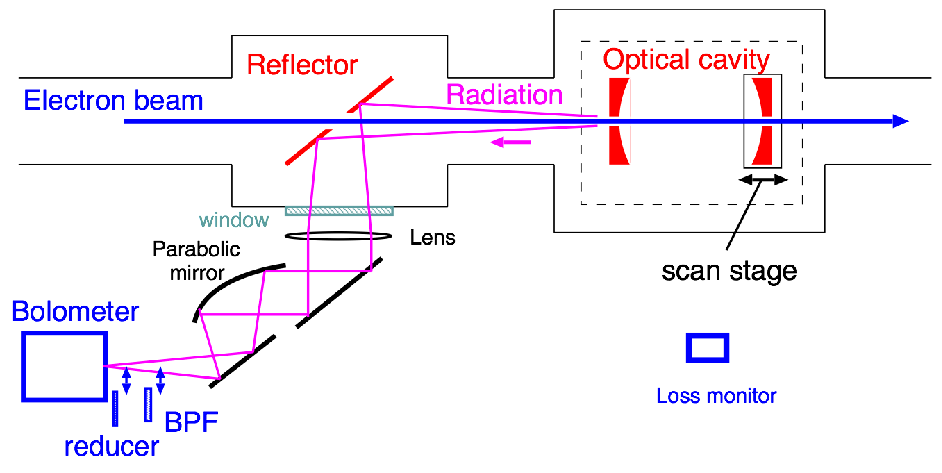}% Here is how to import EPS art
\caption{\label{fig:cavitylayout} The experimental layout. 
The radiation emitted 
toward the upstream direction
was measured. A bolometer was placed at the focal point. 
}
\end{figure}

Figure \ref{fig:cavitylayout} shows the layout of the optical cavity system.
Two identical gold-coated copper mirrors were used as cavity mirrors.
The cavity was designed to satisfy the confocal condition.
Both the curvature radius of the mirror
and the cavity length were 115 mm,
which corresponds to a bunch repetition of 1.3 GHz.
The thickness of the mirrors was 10 mm,
the diameter of the mirrors was 50 mm,
and the diameter of the holes at the center was 3 mm.
The eigenmode size
is calculated to be $w=4.7$ mm on the mirror at 0.5 THz. 
The relative alignment of the mirrors
had 50 $\mu$m precision 
using angular adjusters of the mirror holders 
referring mechanical measurements.
In order to scan the cavity length,
the downstream mirror was mounted on a piezo stage,
which can be controlled up to 1 nm precision within a 20 mm range.
The entire cavity structure was mounted on a manipulator 
so that the structure could be removed from the beam axis in the vacuum chamber.

A scintillator screen can be inserted at the center of the cavity
in order to tune the 
electron beam size and position in the cavity.
To detect local beam loss in the cavity,
beam loss monitors  \cite{lossmon}
were installed near the CDR cavity.

A fraction of the radiation power inside the cavity
was emitted through the holes in
both directions.
Only emission in the upstream direction was measured.
The radiation was reflected along the transverse direction
by a reflector installed in a separate chamber 309 mm upstream of the cavity.
The reflector was a gold-coated flat stainless steel plate mirror
angled 45 degrees with respect to the beam line.
It has an elliptical hole,
which appeares as a 10 mm diameter circular aperture from the perspective of the electron beam.
The radiation was transmitted to air through a sapphire window.

The radiation extracted from the vacuum chamber
was first collimated 
by a lens.
Then, 
it was focused by a parabolic mirror.
At the focal point,
a liquid-helium cooled Si bolometer (Infrared Laboratories, Inc.)
was used as the terahertz radiation detector.
We prepared a band-pass filter 
with 0.5 THz center frequency and 10 \% bandwidth 
(fullwidth half-maximum),
and this filter was inserted in front of the detector.

%%%%%%%%%%%%%%%%%%5
\paragraph{Experimental Result}

\begin{figure}[b]
\includegraphics[width=0.9\linewidth, height=0.45\linewidth]{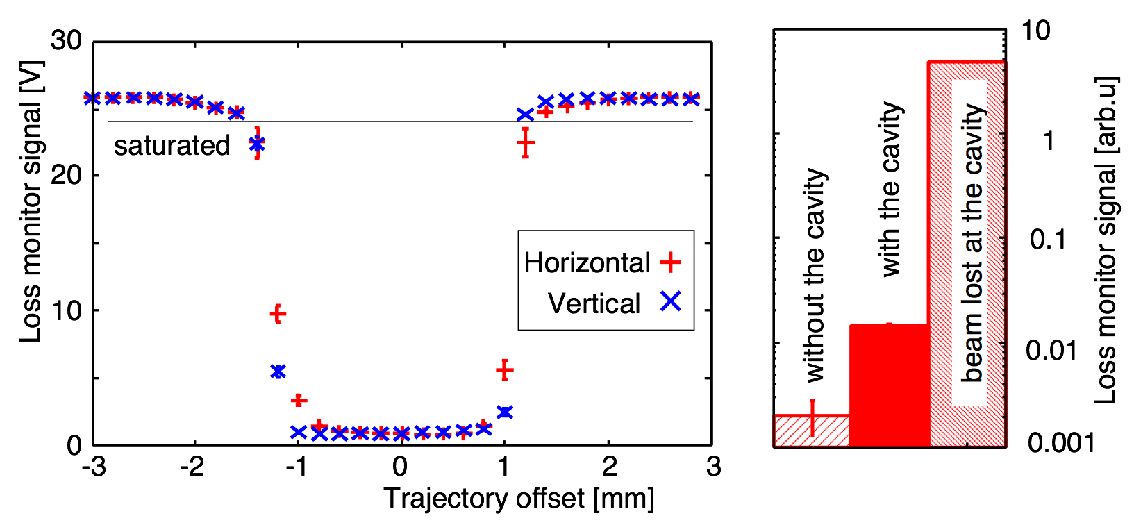}% Here is how to import EPS art
\caption{\label{fig:lossmon} 
(left) Signal strength of the loss monitor as a function of beam position at the cavity.
Note that the loss monitor was saturated when the beam hit at the cavity.
(right) A comparison of three cases was used
 to estimate the beam loss fraction.
In this measurement, 
the gain of the loss monitor was reduced to avoid saturation.}
\end{figure}

Referring to the signal at the beam loss monitors,
the local beam trajectory at the cavity was scanned.
Figure \ref{fig:lossmon} shows the loss monitor signal
as a function of the beam position offset.
It shows that the beam trajectory was well optimized 
to be at the center of the holes
with apploximately  $\pm$1 mm clearance.
A quantitative estimation of the beam loss was performed by
comparing the loss monitor signal in three cases:
without the cavity, when beam passes through the cavity, and when the beam hits at the cavity.
The fraction of beam loss at the cavity 
was estimated to be 2500 ppm.

\begin{figure}[b]
\includegraphics[width=0.9\linewidth, height=0.9\linewidth]{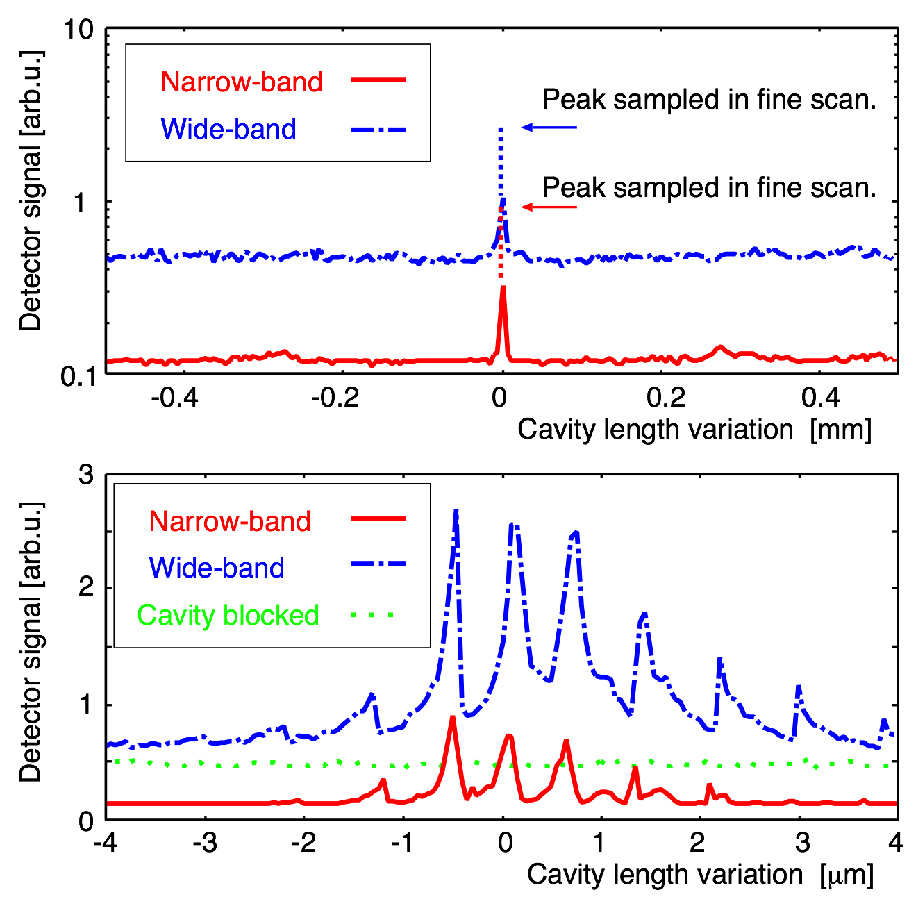}% Here is how to import EPS art
\caption{\label{fig:resonancedata} Cavity length scan results. 
(top) Wide range scans in 5 $\mu$m increments.
Since the peak was not correctly sampled in the rough step,
the actual peak heights are shown with arrows.
(bottom) Fine scans at the peaks in 50 nm increments.
}
\end{figure}

The cavity length was scanned while measuring 
the bolometer signal.
Figure \ref{fig:resonancedata} shows the results
from narrow-band and wide-band measurements,
corresponding to the measurements taken with and without the 0.5 THz band-pass filter, respectively.
A sharp peak was observed, indicating the cavity was excited at resonance.
In the narrow-band measurement,
small peaks were repeatedly observed with 0.3 mm separation
along the cavity length,
which corresponds to a half-wavelength of a 0.5 THz wave.
A precise scan around the peak was performed
to determine the profile of the resonance peak,
It turned out that
the peak was split into fine structures.
The width of a single peak was measured to be $\sim$150 nm.
The overall shape of the peak looks 
similar for both wide-band and narrow-band measurements.
Figure \ref{fig:resonancedata} (bottom) also shows 
wide-band measurement data
when the screen monitor was inserted in the cavity for blocking the cavity resonance.
We confirmed that the resonance peaks disappeared in this case.
Even in this blocked condition,
the bolometer detected some background signal,
which is treated as a baseline signal that is independent of cavity length.

%%%%%%%%%%5
%%%%%%%%%%%

\begin{figure}[b]
\includegraphics[width=0.9\linewidth, height=0.45\linewidth]{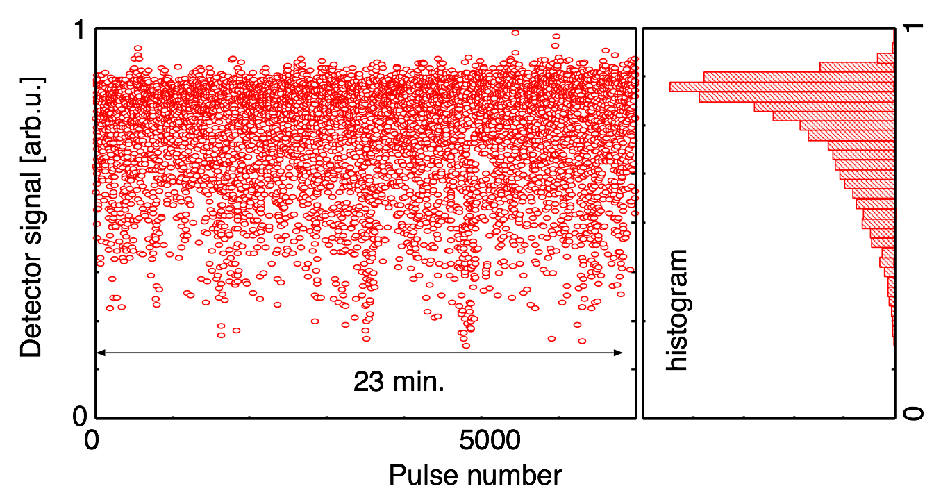}% Here is how to import EPS art
\caption{\label{fig:stability} Long term stability measurement.
The sharp resonance was maintained for 23 min.}
\end{figure}

In order to evaluate the stability of the system,
long-term data was taken 
with the cavity length fixed at one of the peaks.
Figure \ref{fig:stability} shows the bolometer signal
from 7000 continuous beam pulses,
which corresponds to 23 minutes of operation.
Although shot-by-shot fluctuations were observed,
the cavity tended to remain at  resonance.
74 \% of the data were higher than 75 \% of the maximum.

%%%%%%%%
%%%%%%%%

%%%%%%%%%%%%
\paragraph{Discussion}

The appearance of the resonance structure
in the cavity scan
proves that stimulated emission occurs in the cavity.
Resonance occurs
at the perfect synchronization between 
the radiation round-trip rate in the cavity
and the bunch repetition rate.

From the fine-structured peak width,
the measured finesse was estimated to be $\sim$1000.
The observed finesse is more than one order of magnitude higher
than past stimulated radiation experiments
\cite{lihn,shibata_broadband,aryshev_lucx}.

Comparing the wide-band and narrow-band measurement results,
the width of the peaks looks almost the same.
This means that the resonance condition coincides 
through the wide spectrum of the longitudinal modes.
This confirms that the optical cavity is designed to have a
zero-CEP shift as expected.

The fine structures of the peaks are not what we originally expected.
We guess each peak corresponds to 
higher-order transverse modes in the cavity.
These transverse modes should be degenerate
if the cylindrical symmetry of the cavity was perfect.
The relative mirror misalignment sensitively
affects the cylindrical symmetry of the eigenmodes,
and it can split the resonance conditions.

A baseline offset of about 1/5 of the peak
was observed.
We guess that the origin of the baseline
originates from backward emission of diffracted radiation
at the outside surface of the upstream cavity mirror.

The beam loss in the cavity was
estimated to be 2500 ppm.
However, compared with the beam size and the clearance of the mirror aperture,
the loss should be much lower under the assumption of a simple Gaussian distribution.
There might exist a non-Gaussian beam halo in the electron beam \cite{olga}.
By removing the beam halo
using collimators located at the upstream part of the accelerator,
we expect to be able to reduce the beam loss \cite{akagi,obina_cw},
and the system might be compatible with the CW high current operation
\cite{jlab_MW}.

%%%%%%%%%%%%
\paragraph{Conclusion}
\quad

A scheme based on the CDR is attractive
as a unique accelerator-based light source in the terahertz spectral range.
This scheme may be compatible with 
a high power accelerator, such as ERL,
because it does not destroy the electron beam.
The stimulated radiation mechanism can be considered 
in order to greatly enhance the extraction efficiency 
of radiation from the electron beam,

We performed an experiment to show 
the stimulated radiation process with the CDR layout.
An optical cavity with a small beam hole
was installed along the straight beam pass of an ERL test accelerator,
which could provide 
a low emittance and short-bunch beam with a high repetition rate.
A sharp resonance peak was observed
during the cavity length scan,
indicated stimulated radiation
occurs at the perfect synchronization.
Thanks to the CEP-optimized design of an optical cavity,
cavity modes in a wide-band spectrum
were coherently excited simultaneously.

\begin{acknowledgments}

We appreciate the cERL development team %lead by H.R.Sakai
for their support in regard to the beam operation.
This work was partially supported by
JSPS KAKENHI Grant Number 16H05991 and 18H03473,
and by Photon and Quantum Basic Research Coordinated Development Program
from the Ministry of Education, Culture, Sports, Science and Technology, Japan.

\end{acknowledgments}

\nocite{*}

\bibliography{cdrbib}% Produces the bibliography via BibTeX.

\end{document}